# Security Aspects of the Authentication Used in Quantum Cryptography

Jörgen Cederlöf and Jan-Åke Larsson

*Abstract*—Unconditionally secure message authentication is an important part of quantum cryptography (QC). In this correspondence, we analyze security effects of using a key obtained from QC for authentication purposes in later rounds of QC. In particular, the eavesdropper gains partial knowledge on the key in QC that may have an effect on the security of the authentication in the later round. Our initial analysis indicates that this partial knowledge has little effect on the authentication part of the system, in agreement with previous results on the issue. However, when taking the full QC protocol into account, the picture is different. By accessing the quantum channel used in QC, the attacker can change the message to be authenticated. This, together with partial knowledge of the key, does incur a security weakness of the authentication. The underlying reason for this is that the authentication used, which is insensitive to such message changes when the key is unknown, becomes sensitive when used with a partially known key. We suggest a simple solution to this problem, and stress usage of this or an equivalent extra security measure in QC.

## I. INTRODUCTION

Quantum cryptography (QC), or more accurately quantum key growing (QKG), uses properties of quantum mechanical systems to share a secret key between two sites. QKG was first proposed in 1984 [1] and there are several variations on the theme today [2]–[4]. Because there are excellent descriptions of these systems elsewhere (e.g., [4]), we will only outline the generic steps of a QKG algorithm here, and then focus on the authentication used. The security of QKG is based on laws of nature [5]–[7] rather than computational complexity as is usually the case for key-sharing systems [8], and therefore, we will here not assume that there are any bounds to the computational capacity of the attacker.

We will use common-practice terminology and refer to the sender, receiver, and eavesdropper as Alice, Bob, and Eve, respectively. To set up a QKG system Alice and Bob need a "quantum channel" between them where they can send and receive, or share, quantum systems, e.g., "quantum bits" (qubits). One example is an optical fiber carrying single photons with the qubit coded in the photon's polarization, but there are many other possibilities. In a perfect channel, every qubit sent by Alice is received and correctly measured by Bob, and Bob receives no qubits which Alice has not sent. In practice, such channels do not exist. A real-world channel can lose almost all qubits in transit, make Bob think he received qubits never sent by Alice, and modify some of the qubits that do go from Alice to Bob. However, a perfect channel is not needed. As long as the errors are within some limits, QKG will still produce a key that is both shared and secret [4], [9]–[14].

They will also need a classical communication channel. The alternatives include but are not limited to the Internet, the same optical fiber used above, and a network cable parallel to the optical fiber. Often in this context, a simplifying assumption is used that the classical channel can be eavesdropped on, but not be modified by Eve. Unfortunately, unmodifiable channels do not exist in the real world, so message authentication must be used to allow Alice and Bob to detect Eve's modification attempts. To be able to authenticate, Alice and Bob will need a (small) shared secret key to start with.

The purpose of the QKG system is to use the two channels and a small portion of the already shared key to generate new key portion, larger than the one just used. The initial key only needs to be large enough to allow for the first generation sequence, typically to authenticate two messages, one from Alice to Bob and one in the other direction. This will enable the key to grow somewhat (QKG), and will allow for further runs, in which the key will grow even more. A round consists of a number of steps.

1) *Raw key generation:* Use the quantum channel to transmit/generate a bit sequence, shared between Alice and Bob but equal only in a portion of the positions. The size of this portion depends on the protocol used, properties of the channel, and whether Eve is listening on the quantum channel.

2) *Sifting:* Remove most of the bits that do not match by comparing parameters of each use of the quantum channel, the "settings." This will discard noisy bits without sending any information about the value of the bits on the classical channel. A smaller "sifted" key is obtained which is equal for Alice and Bob in a considerably larger portion, the size of which depends on properties of the channel and whether Eve is listening.

3) *Error correction, or key reconciliation [15]:* Perform error correction on the sifted key and estimate the error rate to detect whether Eve was listening on the quantum channel, either with a few sacrificed bits from the sifted key, or with some of the sifted-out bits from the last step, depending on details of the protocol. If the error rate is above a predetermined bound, Alice and Bob conclude that Eve has been listening and the round must be aborted.

4) *Privacy amplification [16]–[18]:* If the noise is lower than the predetermined bound, Eve may still have been listening but in that case she has opted to only extract very little information. In this case, Alice and Bob can perform "privacy amplification" to lower Eve's information even further, sacrificing a few bits of their candidate key in the process.

5) *Authentication [19]–[21]:* The final step of each round is to authenticate the messages sent from Alice to Bob and from Bob to Alice on the classical channel, to make sure Eve has not modified these messages. The sender uses key bits from the previously shared secret key to create an authentication tag from the message. The used key bits are then discarded. The tag is sent along with the message and the recipient uses his copy of the key to generate another tag from the received message. If the tags are identical, the message is accepted as authentic and the new key just generated is added to the remaining key from the last round. If the authentication fails, Eve is assumed to be trying to interfere and the round should be aborted. (A complication is the fact that the error correction is not perfect. An error can, with a small probability, sneak through. If that error is in the key used for authentication in a later round, the authentication will fail even without Eve being present.)

There are variations in the details but all QKG protocols contain these main steps. Eve's presence is detected via high error rate on the quantum channel in step 3) or failure of authentication on the classical channel in step 5). If the authentication step is not performed, all QKG protocols are susceptible to a man-in-the-middle attack, where Eve would impersonate Bob when communicating with Alice and vice versa. Even when performing authentication, one broken round will provide Eve with the authentication key for a subsequent round and can break that too, and so on for all future rounds. We will examine the authentication step of the protocols in some more detail here and show that it is also sensitive to the choice of the message to be authenticated.

## II. AUTHENTICATION

In QKG, the standard is to use Wegman–Carter authentication [19]–[21]. This is the authentication equivalent of the Vernam cipher (the one-time pad; see, e.g., [22]), for which all messages are equally likely if the key is unknown. In Wegman–Carter authentication, all values of the tag are equally likely if the key is unknown, and even if one message–tag pair is known, all values of the tag corresponding to another message still are (almost) equally likely. A tag is shorter than a message, so in comparison, just guessing a tag will be more likely to succeed than the corresponding guess of a message in the Vernam cipher. Nevertheless, given a sufficiently long tag length, the probability of correctly guessing the tag will be very low in Wegman–Carter authentication. That is, the probability of generating the correct tag for a forged message will be very low.

In the Vernam cipher, the required key needs to be at least as long as the message to be encrypted. Fortunately, in Wegman–Carter authentication, the required key grows only logarithmically with the message length. This is essential for QKG as it is then only a matter of making the rounds large enough to gain more key than is lost in the authentication.

Formally, the fundamental building block of Wegman–Carter authentication is called *universal families of hash functions*,[1] a family $\mathcal{H}$ of functions that map a message in the set of possible messages $\mathcal{M}$ to a tag in the set of tags $\mathcal{T}$. The following formal definition of the appropriate family of hash functions is taken from [21].

*Definition 1 ($\epsilon$-almost strongly universal$_2$ ($\epsilon - ASU_2$) hash functions):* Let $\mathcal{M}$ and $\mathcal{T}$ be finite sets and call functions from $\mathcal{M}$ to a tag in the set of tags $\mathcal{T}$. Let $\epsilon$ be a positive real number. A set $\mathcal{H}$ of hash functions is $\epsilon$-almost strongly universal$_2$ if the following two conditions are satisfied:

1) The number of hash functions in $\mathcal{H}$ that takes an arbitrary $m_1 \in \mathcal{M}$ to an arbitrary $t_1 \in \mathcal{T}$ is exactly $|\mathcal{H}|/|\mathcal{T}|$.

[1]A word of warning is perhaps appropriate regarding terminology, as these hash functions are quite different from "cryptographically secure hash functions" sometimes mentioned in connection with authentication. It is impossible to construct unbreakable cryptographically secure hash functions (see, e.g., [23]). They have similarities and both deserve to be called hash functions, but the individual hash functions of Wegman–Carter are not, and need not be, cryptographically secure in the classical sense.

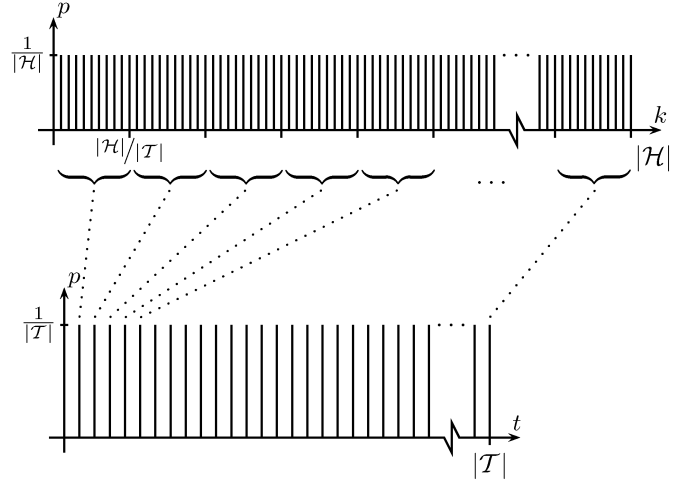

Fig. 1. In Wegman–Carter authentication, a given message $m$ organizes the keys $k$ into subsets that each map the message to one value of the tag $t = h_k(m)$, and these subsets are of equal size (for an $\epsilon$-ASU$_2$ family of hash functions). That is, to Eve, the key $K$ is completely unknown (uniformly distributed), and therefore, so is the tag $T_E = h_K(m_E)$ for her message $m_E$.

2) The fraction of those functions that also takes an arbitrary $m_2 \neq m_1$ in $\mathcal{M}$ to an arbitrary $t_2 \in \mathcal{T}$ (possibly equal to $t_1$) is no more than $\epsilon$.

The parameter $\epsilon$ controls a tradeoff between the size of $\mathcal{H}$ and the probability to guess the correct tag. The lower bound of $\epsilon = 1/|\mathcal{T}|$ can be achieved if a large family can be tolerated, and Wegman and Carter included several such examples in [19]. Those families are too large to be usable in QKG, but Wegman and Carter later showed [20] that by just doubling the possibility of a correct guess, a much smaller $2/|\mathcal{T}|$-ASU$_2$ family can be constructed. That family is small enough for QKG, and although there are many other similar families, the exact choice is not important and we will use their original example from [20].

In formal language, the authentication proceeds as follows. Alice and Bob share a secret key $k$ just large enough to select a hash function $h_k \in \mathcal{H}$, $0 \le k < |\mathcal{H}|$. Alice wants Bob to have the message $m_A \in \mathcal{M}$ and sends both $m_A$ and $t_A = h_k(m_A)$. Bob verifies that $t_A$ really equals $h_k(m_A)$ and accepts the message as authentic if it does. The key $k$ is then discarded and never reused.

Let us now introduce Eve who has control over the channel between Alice and Bob and wants Bob to accept a faked message $m_E \in \mathcal{M}$. To her the secret key is a random variable $K$ uniform over its whole range $0 \le K < |\mathcal{H}|$. If the key is a random variable, so is the tag for her message $T_E = h_K(m_E)$. The first condition of Definition 1 says that if $K$ is uniform over its whole range, so is $T_E$ (see Fig. 1). Eve can take a guess, but any guess $t$ is correct only with the probability

$$P(T_E = t) = 1/|\mathcal{T}|. \tag{1}$$

Eve may also wait until Alice tries to send an authenticated message to Bob, pick up the message and the tag, and make sure Bob never sees them. With both $m_A$ and $t_A = h_K(m_A)$ at her disposal, she can, given enough computing power, rule out all keys that do not match and be left with just $1/|\mathcal{T}|$ of the keys to guess from; see Fig. 2. However, the second condition of Definition 1 says that even with this knowledge, any tag value $t$ guessed by Eve is correct (equal to the correct tag $T_E$) for her $m_E \neq m_A$ (with $K$ uniform over its whole range) at best with the probability

$$P(T_E = t \mid h_K(m_A) = t_A) \le \epsilon. \tag{2}$$

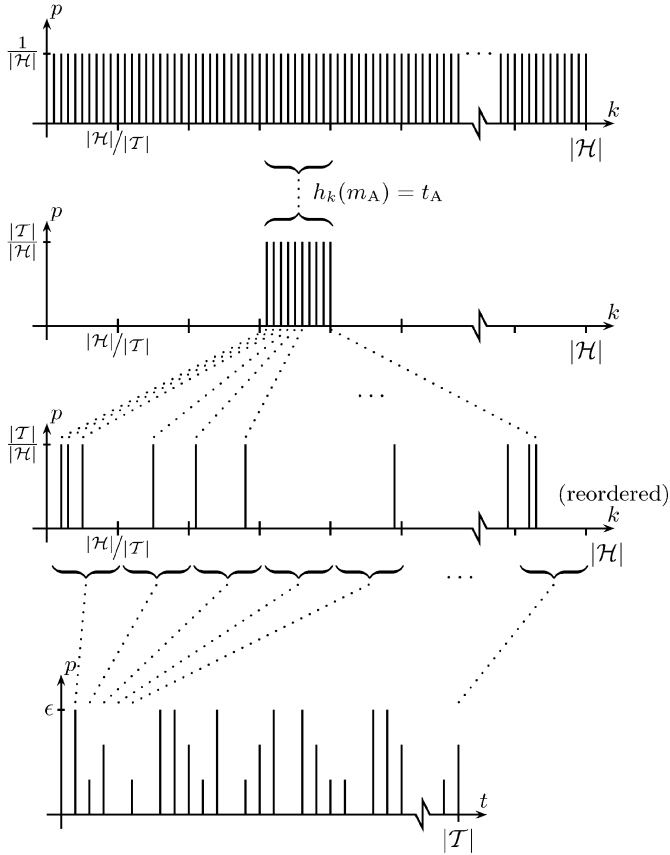

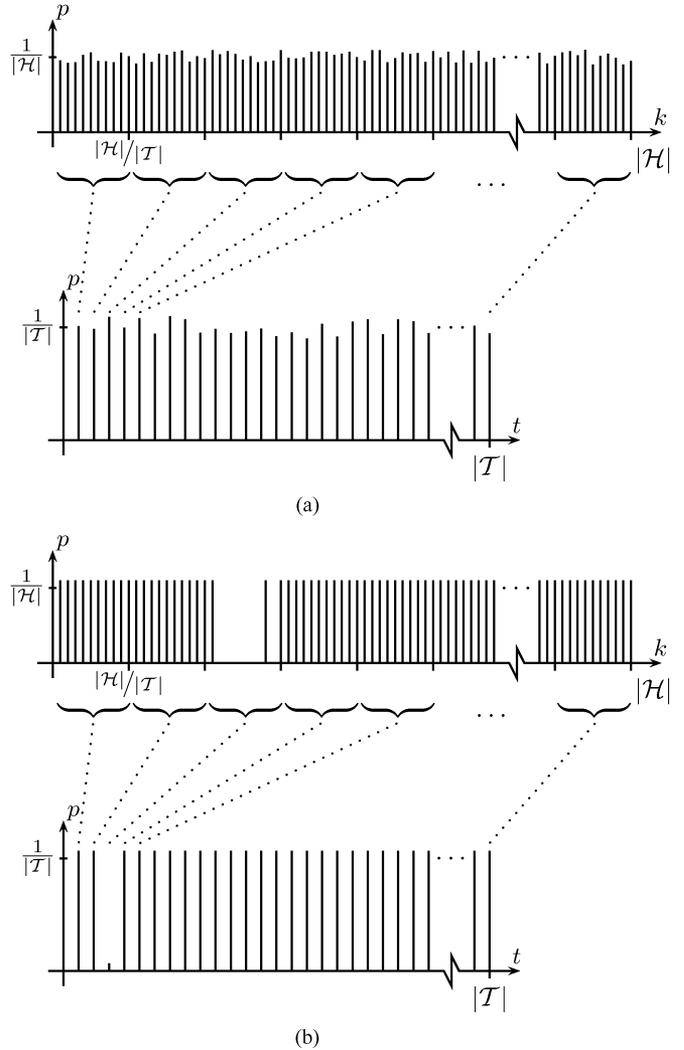

Fig. 2. In Wegman–Carter authentication, a given message–tag pair corresponds to one subset of keys that map the message onto that tag value. A different message induces a different family of subsets, and will spread out the remaining keys so that all tag values have a probability less than or equal to $\epsilon$ (for an $\epsilon$-ASU$_2$ family of hash functions, if the keys are equally probable).

Fig. 3. Eve's information on the key will induce a nonuniform distribution on $k$, and also on $t$. (a) Nonuniform distribution on $k$ induces a nonuniform distribution on $t$. (b) Distribution can be very skew, for instance, if Eve holds information that allows her to rule out some keys entirely.

The parameter $\epsilon$ is clearly an upper limit on the probability that Eve makes the right guess and manages to fool Bob into accepting a fake message, at least if Eve knows nothing about the key beforehand.

In fact, Wegman–Carter authentication is cryptographically secure in the following way: the probability of Eve guessing the tag value for her message $m_E$ does not depend on which message $m_A$ Alice sends, as long as it is not equal to Eve's message $m_E$. The probability is always less than $\epsilon$, independently of $m_A$, or put in other words, there are no message–tag pairs from Alice that are significantly weaker than others. Even if Eve was allowed to choose $m_A$ (different from $m_E$) and was given the tag for that message, she would not be in an improved situation in regards to the tag $T_E$ corresponding to her message $m_E$. This may not seem important at this point, but will prove to be interesting later.

If Eve tries to break the authentication in the above scenario and fails, her presence will be detected and the QKG round will be aborted. A complicating factor is that the authentication can fail from time to time without Eve because of channel noise, so Eve can try to break the authentication, but to avoid raising suspicion, she should only do this seldom. The parameter $\epsilon$ should be chosen so that even if Eve does this, the expected life of the system is long enough for Alice's and Bob's needs. For the $2/|\mathcal{T}|$-ASU$_2$ family from [20], a 32-bit tag would give a probability of $2^{-31}$ to generate the correct tag after having seen a message–tag pair. On average, Eve would need $2^{31} \approx 2.1 \times 10^9$ attempts. If one extra failure of the authentication, e.g., every 10 s, is not detectable, it would take on average 680 years to guess the correct tag. This would be long enough for most uses.

## III. PARTIALLY KNOWN KEY

In the previous section, we have assumed that Eve has no information on the secret key used in the authentication, i.e., to Eve, the key $K$ was a random variable uniform over its whole range. This is an unrealistic requirement in QKG. Information leakage in the quantum transmission phase is unavoidable but the damage can be reduced by using privacy amplification, which will reduce Eve's knowledge of the key significantly, but not all the way to nothing. As soon as the whole preshared key is used, Alice and Bob will have to start trusting authentication with a key that is not completely secret.

If Eve has some information on the key, obtained from earlier rounds of the QKG protocol, but has not seen any message–tag pair (as depicted in Fig. 3), an upper bound for the chance that Eve's generated, or guessed, tag value $t$ is correct is the sum of probabilities for the $|\mathcal{H}|/|\mathcal{T}|$ most probable keys. The appropriate bound for Eve's knowledge on the key is given by the min entropy

$$H_\infty(K) = \min_k (-\log_2 P(K = k)). \tag{3}$$

For a given value of the min entropy, the chance of a correctly guessed tag value is maximized if the probabilities $P(K = k)$ are all equal.

This occurs when Eve uses all her information to eliminate some keys, and we denote the remaining keys

$$\mathcal{H}_E = \mathcal{H} \setminus \{h_1, \ldots, h_n\}. \tag{4}$$

This means that from her perspective the true key is drawn from the remaining $|\mathcal{H}_E| = r|\mathcal{H}|$ keys with equal probability [i.e., $H_\infty(K) = \log_2 r|\mathcal{H}|$; see Fig. 3(b)]. We arrive at

$$P(T_E = t) \leq \sum_{1}^{|\mathcal{H}|/|\mathcal{T}|} \frac{1}{r|\mathcal{H}|} = \frac{1}{r|\mathcal{T}|}. \tag{5}$$

The probability of a correct guess increases, but only a little if the parameter $r$ is close to 1. If Eve knows nothing about the key, her key (min)entropy equals the size of the key, and the probability is (bounded by) $1/|\mathcal{T}|$ as expected.

Now, when Eve has a little knowledge on the key *and* picks up a message–tag pair, she again gains additional information that increases her knowledge about the key. The message–tag pair $m_A + t_A$ that Eve receives from Alice identifies a subset of keys (hash functions) of size $|\mathcal{H}|/|\mathcal{T}|$ from which the key must have been drawn

$$\mathcal{H}_{t_A} = \{h \in \mathcal{H} : h(m_A) = t_A\}. \tag{6}$$

Given that the set of possible keys is $\mathcal{H}_E$ rather than $\mathcal{H}$, the final set of possible keys is not $\mathcal{H}_{t_A}$ but $\mathcal{H}_{t_A} \cap \mathcal{H}_E$. In the extreme case, when only one key remains in this subset, Eve will know which key was used by Alice, and in this case, she can simply create a tag using the identified key. However, it is also possible to use the result if more than one key is present in $\mathcal{H}_{t_A} \cap \mathcal{H}_E$ as in Fig. 4. More specifically, when

$$|\mathcal{H}_{t_A} \cap \mathcal{H}_E| \leq \epsilon |\mathcal{H}|/|\mathcal{T}| \tag{7}$$

there may exist messages $m$ that are such that

$$\forall h_1, h_2 \in \mathcal{H}_{t_A} \cap \mathcal{H}_E, h_1(m) = h_2(m). \tag{8}$$

That is, for this message, all remaining keys map to the same tag. The maximum number $\epsilon |\mathcal{H}|/|\mathcal{T}|$ is given in requirement 2) in Definition 1. The number of messages with this property will increase as $|\mathcal{H}_{t_A} \cap \mathcal{H}_E|$ decreases from $\epsilon |\mathcal{H}|/|\mathcal{T}|$. If one of these messages coincides with $m_E$, Eve can successfully break the authentication. She may not know exactly which key $k$ was drawn but she knows enough to create the correct tag $t_E = h_k(m_E)$ for her message.

Even when her preferred message $m_E$ does not coincide with one of the above messages, Eve has some freedom in choosing $m_E$ and may be able to adjust her message so that she can use the above technique. The worst possible case is when Eve can choose her message $m_E$ so that she can generate the correct tag $t_E$ for it as soon as (7) holds. We will restrict ourselves to deal with this worst case scenario here and assume that Eve is able to do just this; see further comments in Section VI. This assumption also implies that even if $|\mathcal{H}_{t_A} \cap \mathcal{H}_E| > \epsilon |\mathcal{H}|/|\mathcal{T}|$, she can choose her message $m_E$, so that $\epsilon |\mathcal{H}|/|\mathcal{T}|$ of the key values in $\mathcal{H}_{t_A} \cap \mathcal{H}_E$ give the correct tag for her message. The probability of generating the correct tag given these two sources of information is bounded by

$$P(T_E = t | K \in \mathcal{H}_{t_A} \cap \mathcal{H}_E) \leq \frac{\epsilon |\mathcal{H}|/|\mathcal{T}|}{|\mathcal{H}_{t_A} \cap \mathcal{H}_E|}. \tag{9}$$

Before Eve has seen the tag $t_A$, her chance of success is

$$\begin{aligned}
P(T_E = t & \mid K \in \mathcal{H}_E) \\
&= \sum_{\tau=1}^{|\mathcal{T}|} P(K \in \mathcal{H}_\tau \cap \mathcal{H}_E) \times P(T_E = t | K \in \mathcal{H}_\tau \cap \mathcal{H}_E) \\
&\leq \sum_{\tau=1}^{|\mathcal{T}|} \frac{|\mathcal{H}_\tau \cap \mathcal{H}_E|}{r|\mathcal{H}|} \times \frac{\epsilon |\mathcal{H}|/|\mathcal{T}|}{|\mathcal{H}_\tau \cap \mathcal{H}_E|} = \frac{\epsilon}{r}.
\end{aligned} \tag{10}$$

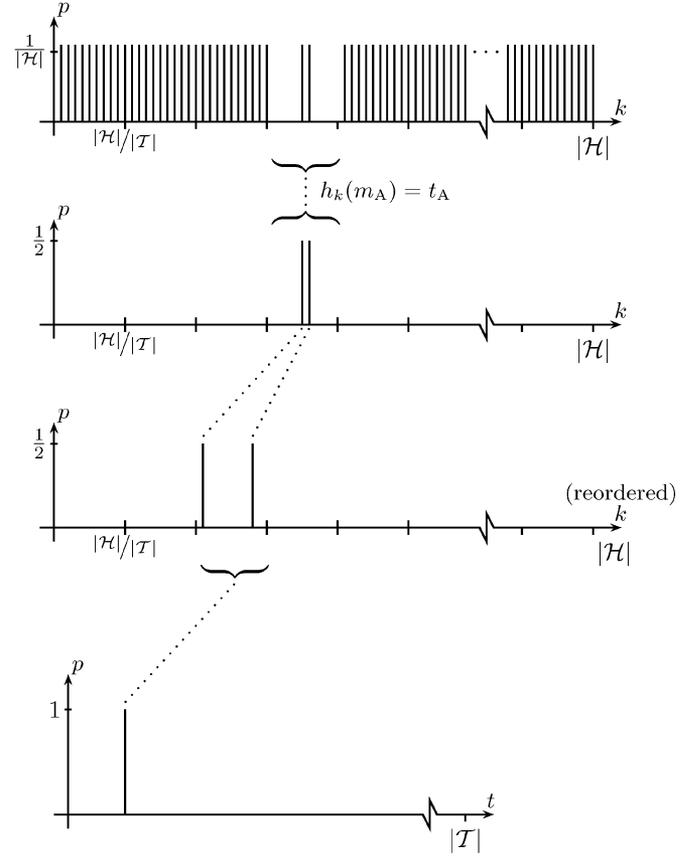

Fig. 4. If Eve can rule out certain keys with her very limited information, it may happen that Alice's message-tag pair allows Eve to rule out all keys except for a few that map her message to the same tag. She can now send her message and that tag, knowing that Bob will accept it. There is no risk whatsoever that Bob will detect her.

The increase in probability from (2) is small if $r$ is close to 1; this suggests that the system is secure (see, also, [24]).

However, if Eve gets to see *both message and tag* before she must decide whether to replace them with her own, the average probability in (10) is not appropriate for comparison with the bound in (2). Instead, the bound in (9) should be used. *But that bound is not a bound:* the right-hand side reaches 1 if there remain at most $\epsilon |\mathcal{H}|/|\mathcal{T}|$ keys in $\mathcal{H}_{t_A} \cap \mathcal{H}_E$. In this situation, Eve has information at hand that enables her to determine whether her attack will be successful, *before* she has replaced the message–tag pair. She may now choose to replace the message–tag pair for her own *only* in those cases when she knows she will be successful, and remain undetected when she is uncertain of success.

The full attack would be as follows: Eve can choose to tap the quantum channel in such a way that the disturbance is below the noise limit set by Alice and Bob. Her aim is not to use the information she gathers to decode messages sent with the generated key, but to break the authentication of the QKG system. She then intercepts each message–tag pair sent by Alice and uses the additional information provided by the pair to determine the tag for her forged message. She will only be successful occasionally, when the following occurs:

1) the message $m_A$ sent by Alice is such that at least one of the subsets depicted in Fig. 4 contain less than $\epsilon |\mathcal{H}|/|\mathcal{T}|$ keys;
2) the key, randomly drawn to Eve, ends up in such a subset.

Because Eve can determine when the attack is successful, i.e., when the remaining keys all map her message to the same tag, she will only replace Alice's message–tag pair on the classical channel when she is certain of success. As long as Eve stays passive she does not risk detection, and she actively replaces the message–tag pair only when her tag is correct. This attack is possible to perform each round, instead of the sparse attempts that the previously mentioned guessing strategy allowed.

In what follows, to simplify the analysis, we will assume that Eve performs the active replacement only when she is certain of success, even though this is not strictly necessary. Eve's probability of success is bounded by (9), and it would be possible to devise a more complicated guessing strategy to be used by Eve when it is less than one, but that is beyond the scope of this correspondence.

## IV. SECURITY?

Let us assess the severity of this threat by estimating the probability that Eve receives the right message–tag pair given only a little information on the key. First, we will also assume that Eve can do nothing more than remove keys essentially at random with her initial knowledge of the key. The message–tag pair that Eve receives corresponds to drawing $|\mathcal{H}|/|\mathcal{T}|$ keys from $\mathcal{H}$ without returning them. The true key will always be present in the drawn keys (and is, of course, one of the remaining possible keys), while the other $|\mathcal{H}|/|\mathcal{T}| - 1$ keys are drawn from $|\mathcal{H}| - 1$ keys of which $r|\mathcal{H}| - 1$ are "possible," i.e., belong to $\mathcal{H}_E$. The number of drawn possible keys $X$ is a random variable, and removing the true key, the random variable $(X - 1)$ will be hypergeometrically distributed

$$(X - 1) \in \mathrm{H_{YP}}\left(|\mathcal{H}| - 1, \frac{|\mathcal{H}|}{|\mathcal{T}|} - 1, \frac{r|\mathcal{H}| - 1}{|\mathcal{H}| - 1}\right). \quad (11)$$

In other words

$$P(X = i) = \frac{\binom{r|\mathcal{H}| - 1}{i - 1}\binom{|\mathcal{H}| - r|\mathcal{H}|}{|\mathcal{H}|/|\mathcal{T}| - i}}{\binom{|\mathcal{H}| - 1}{|\mathcal{H}|/|\mathcal{T}| - 1}}. \quad (12)$$

The interesting case is when the number of keys drawn is less than $\epsilon|\mathcal{H}|/|\mathcal{T}|$, or

$$P\left(X \le \epsilon\frac{|\mathcal{H}|}{|\mathcal{T}|}\right) = \sum_{i=1}^{\epsilon|\mathcal{H}|/|\mathcal{T}|} \frac{\binom{r|\mathcal{H}| - 1}{i - 1}\binom{|\mathcal{H}| - r|\mathcal{H}|}{|\mathcal{H}|/|\mathcal{T}| - i}}{\binom{|\mathcal{H}| - 1}{|\mathcal{H}|/|\mathcal{T}| - 1}}. \quad (13)$$

This probability is complicated to evaluate but can be estimated using the Chebyshev inequality

$$P(|X - \mu| \ge c\sigma) \le 1/c^2 \quad (14)$$

which is rather loose, but generally valid, and will be sufficient for our purposes here. It yields

$$\begin{aligned} P\left(X \le \epsilon\frac{|\mathcal{H}|}{|\mathcal{T}|}\right) &= P\left(\mu - X \ge \mu - \epsilon\frac{|\mathcal{H}|}{|\mathcal{T}|}\right) \\ &\le P\left(|X - \mu| \ge \mu - \epsilon\frac{|\mathcal{H}|}{|\mathcal{T}|}\right) \\ &= P\left(|X - \mu| \ge \frac{\mu - \epsilon\frac{|\mathcal{H}|}{|\mathcal{T}|}}{\sigma}\sigma\right) \\ &\le \frac{\sigma^2}{\left(\mu - \epsilon\frac{|\mathcal{H}|}{|\mathcal{T}|}\right)^2}. \end{aligned} \quad (15)$$

In our case, the mean value is

$$\mu = \left(\frac{|\mathcal{H}|}{|\mathcal{T}|} - 1\right)\frac{r|\mathcal{H}| - 1}{|\mathcal{H}| - 1} + 1 \quad (16)$$

and the standard deviation is

$$\sigma = \sqrt{\left(\frac{|\mathcal{H}|}{|\mathcal{T}|} - 1\right)\frac{r|\mathcal{H}| - 1}{|\mathcal{H}| - 1}\left(1 - \frac{r|\mathcal{H}| - 1}{|\mathcal{H}| - 1}\right)\frac{|\mathcal{H}| - |\mathcal{H}|/|\mathcal{T}|}{|\mathcal{H}| - 2}}. \quad (17)$$

This simplifies considerably in the asymptotic regime

$$r|\mathcal{H}| \gg r|\mathcal{H}|/|\mathcal{T}| \gg 1 \quad (18)$$

where we have

$$\mu = r\frac{|\mathcal{H}|}{|\mathcal{T}|} \quad \text{and} \quad \sigma = \sqrt{r(1-r)\frac{|\mathcal{H}|}{|\mathcal{T}|}} \quad (19)$$

which means that

$$P\left(X \le \epsilon\frac{|\mathcal{H}|}{|\mathcal{T}|}\right) \le \frac{r(1-r)\frac{|\mathcal{H}|}{|\mathcal{T}|}}{\left(r\frac{|\mathcal{H}|}{|\mathcal{T}|} - \epsilon\frac{|\mathcal{H}|}{|\mathcal{T}|}\right)^2} = \frac{r(1-r)|\mathcal{T}|}{(r - \epsilon)^2|\mathcal{H}|}. \quad (20)$$

Further, when $r \gg \epsilon$, this simplifies to

$$P\left(X \le \epsilon\frac{|\mathcal{H}|}{|\mathcal{T}|}\right) \le \frac{1 - r}{r}\frac{|\mathcal{T}|}{|\mathcal{H}|}. \quad (21)$$

In practice, the right-hand constant is very small. The $2/|\mathcal{T}|$-ASU$_2$ hash family from [20] is of size

$$|\mathcal{H}| = |\mathcal{T}|^{4\log\log|\mathcal{M}|} \quad (22)$$

e.g., for a 100-kbit message and a 32-bit tag, this translates to

$$|\mathcal{H}| \approx 2^{32 \times 4 \times 17} = 2^{2176} \quad (23)$$

i.e., roughly 2 kbit of key used. If Eve is allowed to have, e.g., 1/8-bit initial knowledge of the key (so that $r \approx 0.917$), her chance to break the system without fearing detection is less than $3.5 \times 10^{-647}$ each round. At 1000 rounds/s, Eve's expected time to break the system would be at least $10^{635}$ years, much longer than when just guessing once every 10 s. Remember that using this approach, Eve does not guess the tag value but only tries to break the system when she is certain of success. Again, this seems to suggest the same as above; even if Eve has a little information on the key used for authentication, her chances at breaking the authentication do not increase substantially. However, Eve can do more than just wait for the right message–tag pair to arrive; she may have a cunning plan.

## V. POSSIBLE ATTACK

Eve's main obstacle above is the Chebyshev inequality. Viewed in another manner, the central limit theorem ensures that most of the subsets will, with high probability, contain a number of remaining keys very close to $r|\mathcal{H}|/|\mathcal{T}| \gg \epsilon|\mathcal{H}|/|\mathcal{T}|$. Eve's chances of breaking the authentication would increase dramatically if the remaining keys were split into subsets of only two kinds: with either $\epsilon|\mathcal{H}|/|\mathcal{T}|$ or $|\mathcal{H}|/|\mathcal{T}|$ keys in each subset. This will change the probability distribution discussed above, so that the argument that used the Chebyshev inequality does not apply anymore. Eve would then be able to break the authentication if the correct key would happen to fall in a subset with $\epsilon|\mathcal{H}|/|\mathcal{T}|$ remaining keys, since we assume that Eve has enough freedom to generate a message–tag pair of her own as soon as this happens.

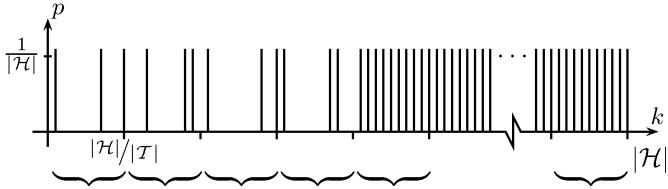

Fig. 5. Eve may be able to influence the message from Alice to arrange for subsets of two kinds, either with $\epsilon|\mathcal{H}|/|\mathcal{T}|$ remaining key values (on the left in the figure) or $|\mathcal{H}|/|\mathcal{T}|$ remaining key values (on the right), to have as many subsets as possible with $\epsilon|\mathcal{H}|/|\mathcal{T}|$ remaining key values.

There are a few methods that Eve could use to arrange the subsets to her liking, but the easiest method would be to change the message: the message from Alice to Bob contains a lot of data that describes what has happened on the quantum channel. Eve can access *and change* what happens on the quantum channel. In essence, Eve has some influence on the content of the message that Alice sends, and as a consequence, Eve can change the subsets. Note that this attack would use a different type of changes on the quantum channel than those caused by Eve extracting information from it, and need not be detectable as an increased noise level in the reconciliation step of the protocol. The attack is different in its aim since it is not intended to increase Eve's information on the key, but rather to maximize the usefulness of the information she has obtained in a previous round. Assuming that Eve does this as best as she can, the subsets may well be such that there remains either $\epsilon|\mathcal{H}|/|\mathcal{T}|$ or $|\mathcal{H}|/|\mathcal{T}|$ keys in each subset (see Fig. 5).

In this situation, the probability of success is instead the probability that the correct key ends up in one of the subsets with $\epsilon|\mathcal{H}|/|\mathcal{T}|$ remaining keys in it. The number of such subsets are

$$
\begin{aligned}
n &= \frac{\sharp \text{ eliminated keys}}{\sharp \text{ eliminated keys in a "good" subset}} \\
&= \frac{(1-r)|\mathcal{H}|}{(1-\epsilon)|\mathcal{H}|/|\mathcal{T}|}
\end{aligned}
\tag{24}
$$

and the probability of ending up in such a subset is

$$
\begin{aligned}
P\left(X \le \epsilon \frac{|\mathcal{H}|}{|\mathcal{T}|}\right) &= \frac{\sharp \text{ possible keys in "good" subsets}}{\sharp \text{ possible keys}} \\
&= \frac{n\epsilon|\mathcal{H}|/|\mathcal{T}|}{r|\mathcal{H}|} = \frac{1-r}{r}\frac{\epsilon}{1-\epsilon}.
\end{aligned}
\tag{25}
$$

The change in probability distribution gives a dramatic increase in probability from the bound in (21) to the value in (25). The difference between $|\mathcal{T}|/|\mathcal{H}|$ and $\epsilon/(1-\epsilon)$ is immense for our $2/|\mathcal{T}|$-ASU$_2$ hash family, since (22) gives

$$
\frac{|\mathcal{T}|}{|\mathcal{H}|} = \frac{1}{|\mathcal{T}|^{4\log\log|\mathcal{M}|-1}} \ll \frac{2}{|\mathcal{T}|} = \epsilon < \frac{\epsilon}{1-\epsilon}.
\tag{26}
$$

In our example, using a $2/|\mathcal{T}|$-ASU$_2$ hash family, a 32-bit tag and 1/8-bit initial knowledge of the key (so that $r \approx 0.9170$), the probability of success is $\approx 4.2 * 10^{-11}$. Again, at 1000 rounds/s, Eve's expected time to break the system would be just *nine months*—nine months to break that QKG system without risk of detection. The immense difference between the two expected times above suggests that this is a problem even when Eve is not able to obtain the ideal subsets.

The real theoretical reason for the existence of this attack is that Wegman–Carter authentication with a partially known key is not cryptographically secure in the way discussed in Section II, concerning Wegman–Carter authentication with a completely secret key. Here, the probability of Eve guessing the tag value for her message $m_E$ does depend on which message $m_A$ Alice sends (even when it is not equal to

Eve's message $m_E$). In other words, *there are message–tag pairs from Alice that are weaker than others*. In QKG, Eve can influence $m_A$ via the quantum channel and is given the tag for that message, and this will improve her situation in regards to the determination of the correct tag $t_E$ corresponding to her message $m_E$.

It is clear that simply sending the tag along with the message to prove authenticity does not work in the long run if Eve has a small but nonzero knowledge of the authentication key used and can influence the message Alice wants to send. The little information carried by the tag can be enough together with what Eve already has, to make Eve certain that her attack will be successful. The probability of this happening in a run is small but Eve can wait, not trying to break the authentication until she is sure of success.

## VI. PREVENTION

To prevent Eve from breaking the QKG system, Alice and Bob may adjust the parameter choices of $|\mathcal{T}|$ and thereby $\epsilon$ by using a larger tag, or $r$ by requiring more privacy amplification. The intent is to decrease the probability in (25), i.e., to make the expected time-of-life of the system long enough to suit their taste. Doing this will use up more key in the authentication, and/or require them to sacrifice more key during privacy amplification. The key production rate of such a system will be lowered, and given the meager output of the systems used today, this is probably not desirable. Minimizing this effect would require a detailed analysis of each individual QKG protocol.

A simpler, more efficient, and generic fix would be to delay the second transfer of information to Eve so that she has to make the decision to try to break the authentication before she knows if she will succeed, i.e., before she has received the tag. The most obvious way to do this is to force Eve to send the message (Alice's or her own) to Bob before she gets hold of the tag.

One solution is using synchronized clocks and sending messages and tags at preagreed times, with a pause longer than the precisions of the clocks. Synchronized clocks are already recommended for other security purposes in present QKG systems; problems with this approach are discussed in [11].

Another and in our opinion better solution that does not need clocks is the following.

1) Alice sends her message $m_A$; Bob receives a message $m$.
2) Bob draws and sends a "salt" $s_B$, a random number drawn from a set at least as large as the set of hash functions $0 \le s_B < |\mathcal{T}|$; Alice receives a salt $s$.
3) Alice calculates a tag based on the concatenation of her message and the received salt $m_A + s$ and sends that tag $t_A = h_k(m_A + s)$; Bob receives a tag $t$ and checks the authentication by comparing $t$ and $h_k(m + s_B)$.

The length of the salt should be at least the tag length because it should be equally difficult to guess as the tag. This would increase the "message length" of the concatenation that is used in the tag generation, but the effect is negligible because the original message is much longer than the tag, and the key used increases logarithmically with the "message length."

When faced with this situation, Eve must decide whether to attack without knowing if she will be successful. If she does decide to attack, there are two ways to proceed:

1) to directly send her message $m_E$ to Bob and either $s_B$ or a faked salt $s_E$ to Alice;
2) to delay sending her message to Bob and send a faked salt $s_E$ to Alice. This will allow her to adjust her message $m_E$ before she sends it to Bob.

Note that, in both cases 1) and 2), Eve needs to actively replace the message and/or the salt on the classical channel *before* she receives the tag from Alice—the tag that carries the extra information Eve needs to determine if her attack will be successful. In this situation, the expression in (10) is the proper bound, and we have restored security.

We mentioned earlier in this correspondence that we assume the worst case scenario where Eve is able to break the authentication provided just that (7) holds. A full analysis of this would necessarily incorporate details of the QKG protocol, including properties of the $\epsilon$-ASU$_2$ family, but we note that the countermeasure presented here, using a salt, is very simple and generic and reestablishes security without the need for such an intricate analysis of each individual QKG protocol.

## VII. CONCLUSION

To conclude, even though Wegman–Carter authentication seems secure when used with a partially known key (see, also, [24]), the usual implementation of a QKG system contains an additional subtlety. Eve can influence the message to be sent, and together with partial knowledge of the key, this opens up Eve's possibilities. Fortunately, a simple remedy exists: force Eve to make her attack before she knows that it will succeed, by making sure Alice will not send the authentication tag until either Bob has received the message or Eve has attempted breaking the system. A real-world implementation of a QKG system might also make it difficult for Eve because 1) Eve's freedom to change the messages to be authenticated might be too limited, and 2) a round normally consists of a dialogue of several messages and an authentication tag for all of them at the very end of the round. Whether this is enough to keep the system secure depends on the details of the system, but implementing the solution proposed here is cheap and requires no deep analysis of the system. We would, therefore, recommend doing just that in future QKG systems.